\documentclass[12pt,twoside,a4paper]{article}
\usepackage{graphicx,lipsum,float,enumerate,caption,framed,emptypage}
\usepackage[authoryear]{natbib}

\usepackage[authoryear]{natbib}
\usepackage{amsmath}
\usepackage{amsfonts}
\usepackage{amssymb}
\usepackage[scale=0.80]{geometry}
\title{Highlights of chaos research}
\author{George Contopoulos\\ Academy of Athens\\ \small gcontop@academyofathens.gr}

\date{}

\begin{document}
\maketitle
\begin{center}\textbf{Abstract}\end{center}

 We describe some highlights in the theory of chaos, that started with Poincare (1899). Generic systems have both ordered and chaotic domains. Chaos appears mainly near unstable periodic orbits. Large chaotic domains are due to resonance overlap. Two recent developments of the theory of chaos, refer (a) to the analytical formulae (Moser series) in the chaotic domains near unstable periodic orbits, and (b) quantum chaos, described by the Bohmian theory of orbits.

\section{1. Introduction}

The father of chaos theory was undoubtedly Poincar\'e (1899). In his ``Me'thodes Nouvelles de la M\'ecanique C\'eleste Vol. III'' discussed the formation of chaotic webs by intersections of the asymptotic curves of unstable periodic orbits. He stated explicitly ``One is impressed by the complexity of this figure, that I will not even try to draw.''

In fact only the use of modern computers allowed us to appreciate the complexity of these curves.

In later years a theory of chaos was developed by mathematicians that considered ergodic systems, i.e. systems whose orbits fill densely the available phase space. A typical example was Birkhoff's book ``Dynamical Systems'' (Birkhoff 1927). For several decades it was considered that generic dynamical systems are either integrable or ergodic. The most important ergodic systems are called mixing, Kolmogorov, or Anosov (Lichtenberg and Liebermann 1992). In particular the first edition of the classical book ``Mechanics'' of Landau and Lifschitz (1976) separates all dynamical systems into two classes, integrable and ergodic.

However it was later realized that in general the dynamical systems contain both order and chaos. Thus in later editions of the book of Landau and Lifschitz the relevant part dealing with integrable and ergodic systems was omitted. 

In fact, integrable and ergodic systems are rather exceptional. Nevertheless they are very important, because they dictate the behaviour of nearby systems. E.g. consider the case of two harmonic oscillators with Hamiltonian:

\begin{equation}
\label{eq:1}
 H_{0}=\frac{1}{2} (\dot{x}^2 + \omega_{1}^2 x^2) + \frac{1}{2} (\dot{y}^2 + \omega_{1}^2 y^2)
\end{equation}

The orbits in this case are in general Lissajous figures, except if the ratio $\omega_{1} / \omega_{2}$, is rational $m/n$ in which case all the orbits are periodic. The case of the Hamiltonian (\ref{eq:1}) is integrable, because in this case there are two integrals of motion, namely $H = const$ and

\begin{align}
\Phi_{0}=\frac{1}{2}(\dot{x}^2 + \omega_{1}^2 x^2)=const\label{eq:2}
\end{align}

Now if we add higher order terms in the Hamiltonian (\ref{eq:1})

\begin{equation}
\label{eq:3}
 H = H_{0} + \epsilon H_{1}
\end{equation}

we can find approximate integrals of the form:

\begin{equation}
\label{eq:4}
 \Phi = \Phi_{0} + \epsilon \Phi_{1} + ...
\end{equation}

Such integrals were found by Whittaker (1916), Cherry (1924, 1926), Birkhoff (1927) and Contopoulos (1960, 1966). In particular in galactic dynamics an integral like (\ref{eq:4}) is called a ``third integral'' because it appears in axisymmetric time independant systems that have already two integrals, the energy $H$ and the angular momentum. Although the third integral is not convergent nevertheless in many cases its truncations provide good approximations of the real orbits. For this reason there are thousands of papers containing applications of the third integral to various problems of celestial mechanics, galactic dynamics, cosmology and also chemical papers on the structure of molecules etc.

\begin{figure}[h]
\centering
\includegraphics[scale=0.35]{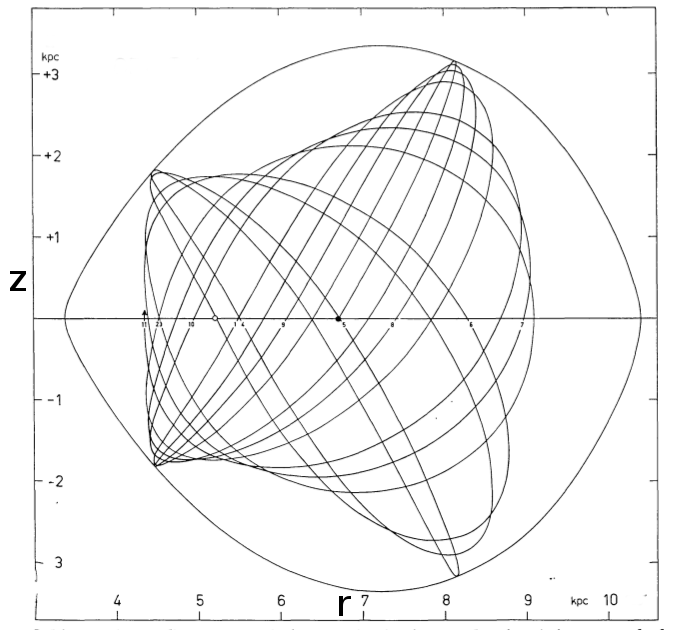}
\caption{A box orbit in the meridian plane of an axisymmetric galaxy (Ollongren 1962).}
\label{fig1}
\end{figure}

When the perturbation $\epsilon$ is small many orbits in a system like (\ref{eq:3}) are close to Lissajous figures (Fig. 1). However there are also tube orbits close to periodic orbits with frequencies $m$ and $n$ (Fig. 2). All these orbits are called ``ordered''. As it was shown by Kolmogorov (1954), Arnold (1961) and Moser (1962) such orbits lie on exact integral surfaces, although the system  (\ref{eq:3}) is nonintegrable. The theorem of Kolmogorov, Arnold and Moser (KAM) was a most important advance in the theory of dynamical systems. In fact the nonconvergence of the formal integrals of the form (\ref{eq:4}) did not allow one to suspect the existence of exact invariant surfaces. In particular Birkhoff (1927) who was one of the pioneers of the dynamical systems, believed that although there are linearly stable periodic orbits in generic dynamical systems, nevertheless these orbits are in fact unstable and nearby orbits escape gradually to large distances.

\begin{figure}[h]
\centering
\includegraphics[scale=0.6]{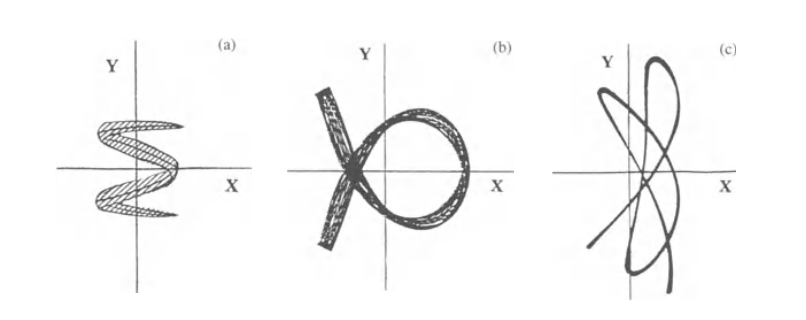}
\caption{Various forms of tube orbits}
\label{fig2}
\end{figure}

On a Poincar\'e surface of section (e.g. $y=0$ in the case of the system (\ref{eq:3}) ) the Lissajous orbits form closed invariant curves around a central point O, while the tube orbits form islands (Fig. 3). The islands surround a set of $n$ points, that represent stable periodic orbits of period $n$ (of period 3 in Fig. 3). However between the islands there are unstable periodic orbits of period $n$ (Fig.3) and near these orbits, the orbits are chaotic.

In fact chaos is introduced mainly nearly unstable periodic orbits. These periodic orbits have one real eigenvalue $\lambda > 1$ and asymptotic curves that are quite irregular. Two nearby orbits starting along an unstable asymptotic curve deviate exponentially in time. If $\Delta s_{0}$ is their initial distance, at the $n$-th intersection with the surface of section their distance along the asymptotic curve becomes approximately $\Delta s= \Delta s_{0} \lambda^n$.\\

In a similar way two nearby orbits in a chaotic domain that are initially at a distance $\Delta s_{0}$ they deviate by a distance $\Delta s = \Delta s_{0} \lambda^t$, where $\lambda$ is a positive number greater than 1. This properly is called ``sensitive dependance on initial conditions'' and it is the main characteristic of chaos.

The successive intersections of a nonresonant orbit by the surface of section form angles $\phi_{i}$, as seen from the central point $O$ and in the limit they define a ``rotation number''
\begin{equation}
    rot \underset{n \to \infty}{=} \frac{1}{n}\sum_{i=1}^{n} \phi_{i}
\end{equation}
The values of $rot$ along a line (say the $x$-axis) forms a rotation curve (Fig.4). At every rational number $rot=m/n$ there is at least one unstable periodic orbit of period $n$ and chaos is formed around it. In fact the KAM theorem applies only in cases where $rot$ is irrational (in fact $rot$ should be ``far from the rationals'' in some sense). In the exceptional cases where $rot$ is rational, or sufficiently close to a rational, there is chaos and (possibly) islands of stability of order $n$. The set of values of $rot$ close to rational is small if the perturbation $\epsilon$ is small, thus in the majority of cases the values of $rot$ are ``far from resonances'' and the orbits form the corresponding KAM curves.

This means that although the rotation curve of Fig. 4 has an infinity of discontinuities (it is called a ``devil's staircase''), nevertheless the important islands of stability where $rot = m/n = const$, are appreciable only for small $n$.

For small $\epsilon$ the various chaotic regions around unstable periodic orbits are separated. But as $\epsilon$ increases the various chaotic regions increase and we have a large degree of chaos due to overlapping of references (Rosenbluth et al. 1966, Contopoulos 1966). Chirikov (1979) studied in detail the resonance overlap phenomenon and for this reason it is called sometimes Chirikov's criterion \footnote{However Chirikov refers to the previous work of Rosenbluth et al. and of Contopoulos, (Zaslavsky and Chirikov 1972).}

\begin{figure}[H]
\centering
\includegraphics[scale=0.7]{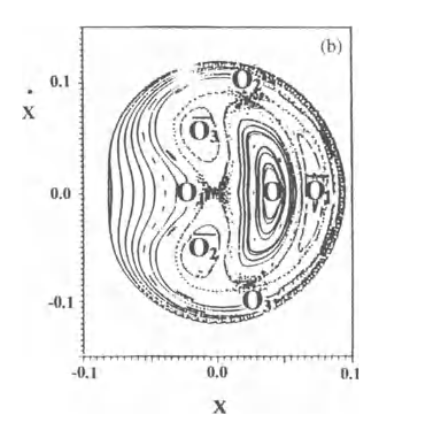}
\caption{Three islands $\bar{O}_1$, $\bar{O}_2$, $\bar{O}_3$ corresponding to tube orbits. Between these islands there is an unstable periodic orbit $(O_1, O_2, O_3)$ surrounded by chaos.}
\label{fig3}
\end{figure}

\begin{figure}[H]
\centering
\includegraphics[scale=0.5]{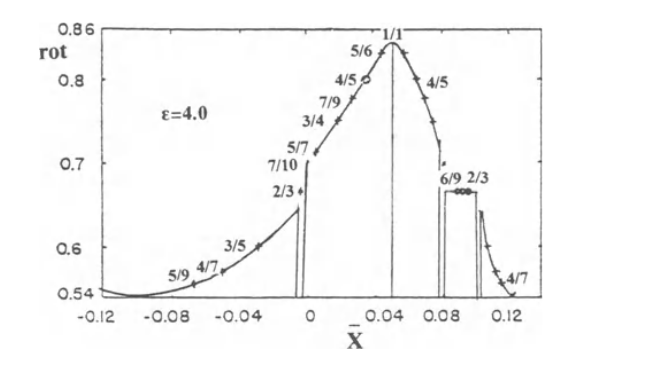}
\caption{A rotation curve giving the rotation number as a function of $\bar{x}$. The curve has small plateaus at every stable periodic orbit (like the plateau $2/3$ on the right) and discontinuities at every unstable periodic orbit (like the discontinuity $2/3$ on the left).}
\label{fig4}
\end{figure}

The phenomenon of resonance overlap is rather abrupt. Namely in a system like that of Fig. 5  we have two major sets of islands of multiplications 2 and 3. For a relatively small perturbation $\epsilon$ although there are chaotic domains near the unstable points of period 2 and 3, these are well separated because the two resonances are well separated by invariant curves around the center O. However as $\epsilon $ increases these separating invariant curves are destroyed and the chaotic domains communicate and chaos increases abruptly.
\begin{figure}[H]
\centering
\includegraphics[scale=0.4]{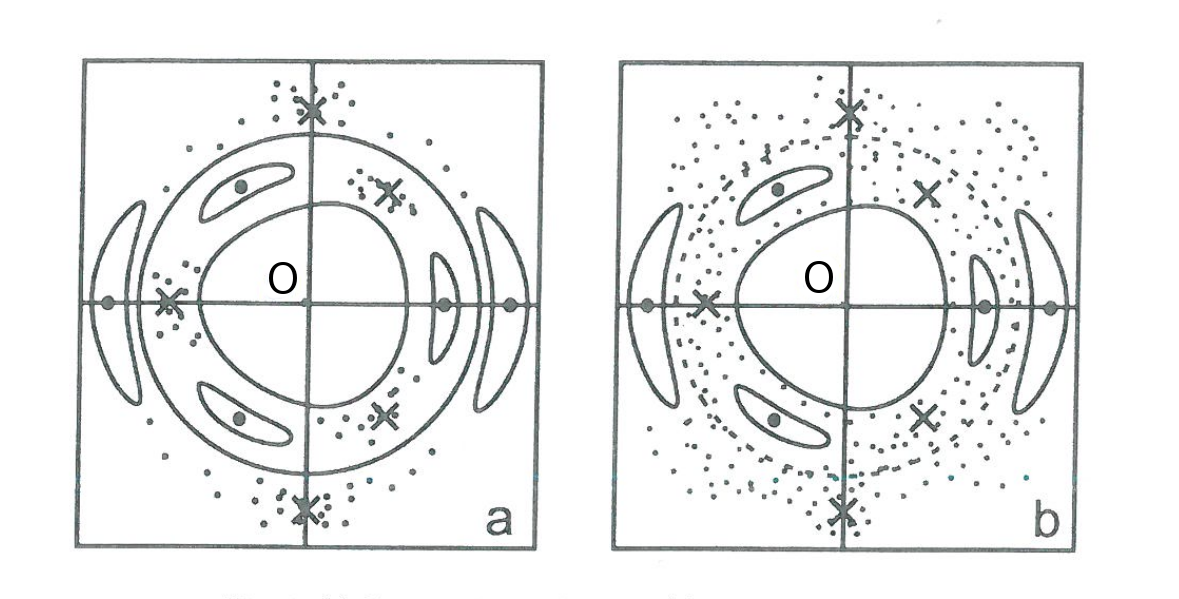}
\caption{(a) The chaotic regions around the double and triple periodic orbits are separated by invariant curves around $O$. (b) When these invariant curves are destroyed they develop an infinity of holes and the chaotic regions overlap.}
\label{fig5}
\end{figure}
In this way we explain the abrupt increase of chaos discoverd by Henon and Heiles (1964) (Fig. 6).
\begin{figure}[H]
\centering
\includegraphics[scale=0.6]{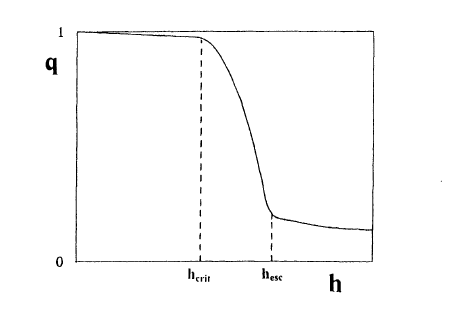}
\caption{The area covered by invariant curves as a function of the energy $h$. This area decreases abruptly at every important resonance $h_{crit}$. Beyond the escape energy $h_{esc}$ there remain islands whose size decreases slowly.}
\label{fig6}
\end{figure}
However the large chaotic domain does not cover the whole available phase space. One finds islands of stability that survive for large values of $\epsilon$ (Fig. 6) (Contopoulos and Polymilis 1987).

These islands are destroyed by successive period doubling bifurcations (Feigenbaum 1978, Coullet and Tresser 1978). In fact at a critical value of $\epsilon=\epsilon_{crit}$ a particular periodic orbit becomes unstable and a double period stable periodic orbit is generated (Fig. 7) producing two smaller islands around the two points of the periodic orbit. The double periodic orbit also becomess unstable for larger $\epsilon_{crit}$ and so on. The intervals between successive bifurcations decrease approximately geometrically and the limit
\begin{align}
\delta=\frac{\epsilon_{crit}(n)-\epsilon_{crit}(n-1)}{\epsilon_{crit}(n+1)-\epsilon_{crit}(n)}
\end{align}
is a universal number. In dissipative systems it was found that $\delta=4.67$ (Feigenbaum 1978, Coullet and Tresser 1978), while in conservative systems $\delta=8.72$ (Benettin et al. 1980, Bountis 1981, Greene and Percival 1981). A different sequence of an infinity of bifurcations due to transitions from stability to instability and vice-versa along the same family of periodic orbits was found numerically by Contopoulos and Zikides (1980) (see also Churchill et al. 1979). In this case however the value of $\delta$ depends on the particular system considered (Heggie 1993).
\begin{figure}[H]
\centering
\includegraphics[scale=0.6]{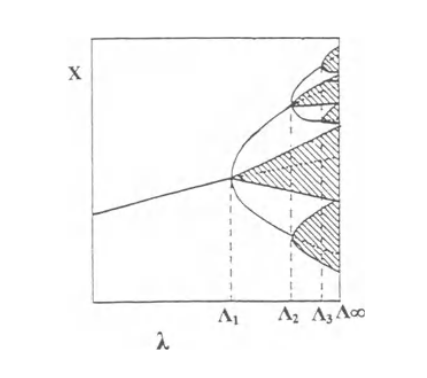}
\caption{Successive period doubling bifurcations and chaotic domains (dark) in a conservative system (the positions $x$ of the periodic orbits are given as functions of a parameter $\lambda$).}
\label{fig7}
\end{figure}
When all the bifurcated orbits become unstable (we have an infinity of unstable periodic orbits)  all the islands produced by bifurcation are destroyed and chaos is dominant. However  for larger $\epsilon$ new islands are formed because of the formation of pairs of stable-unstable periodic orbits by a tangent bifurcation (Contopoulos 1970, 2002). Thus although there are values of $\epsilon$ for which chaos is complete and no islands of stability appear at all, there is no  limiting value of $\epsilon$ beyond which chaos is complete (Contopoulos et al. 2005, Giorgilli and Lazutkin 2000).

A well known case of complete chaos is the cat map (Arnold and Avez 1968) Fig. (8).

\begin{figure}[H]
\centering
\includegraphics[scale=0.5]{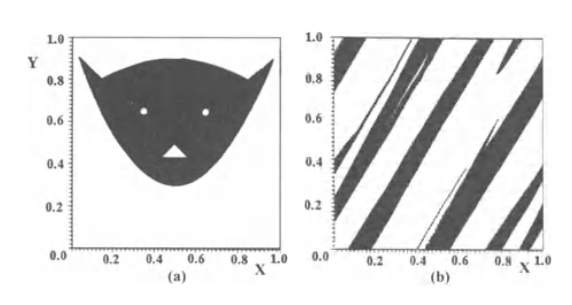}
\caption{The cat map. The original figure (a) is mapped into (b) after two iterations.}
\label{fig8}
\end{figure}

\begin{align}
x'=x+y,\quad y'=x+2y \quad (mod1)
\end{align}
This is a linear system but it is ergodic, mixing, Kolmogorov and Anosov, because of the modulo 1 (otherwise all its orbits would extend to infinity). Similar chaotic systems are the baker map (Lichtenberg and Lieberman 1991) and the stadium (Bunimovich and Sinai 1980). However all these systems have either abrupt returns or abrupt reflections (as in the case of the stadium). If we make these returns or reflections smooth by using a potential we find in general islands of stability. 

As a conclusion we can find Hamiltonian systems with complete chaos (no islands of stability) of the form (3) for particular (isolated) large values of the perturbation $\epsilon$. However between any two chaotic systems with $\epsilon=\epsilon_1$ and $\epsilon=\epsilon_2$ there are systems with intermediate $\epsilon$ where we have islands of stability.

The degree of chaos is measured by a quantity called ``maximal Lyapunov characteristic number'' (LCN):
\begin{align}
LCN=\lim_{n\to\infty}\frac{\ln(\frac{\xi_n}{\xi_0})}{n}
\end{align}
where $\xi_n$ is the nth iterate of an initial deviation $\xi_0$ for a given orbit.

In the case of ordered orbits the Lyapunov characteristic number is zero, while in the case of chaotic orbits it is positive. In practice one calculates the quantity $\chi(t)=\frac{\ln(\frac{\xi}{\xi_0})}{t}$ (finite time LCN) as a function of $t$ (Fig. 9). After a transient time the value of $\log(\chi(t))$ decreases approximately linearly with the logarithm of the time in the case of ordered orbits, while it stabilizes at a constant positive value in the case of chaotic orbits.

\begin{figure}[H]
\centering
\includegraphics[scale=0.3]{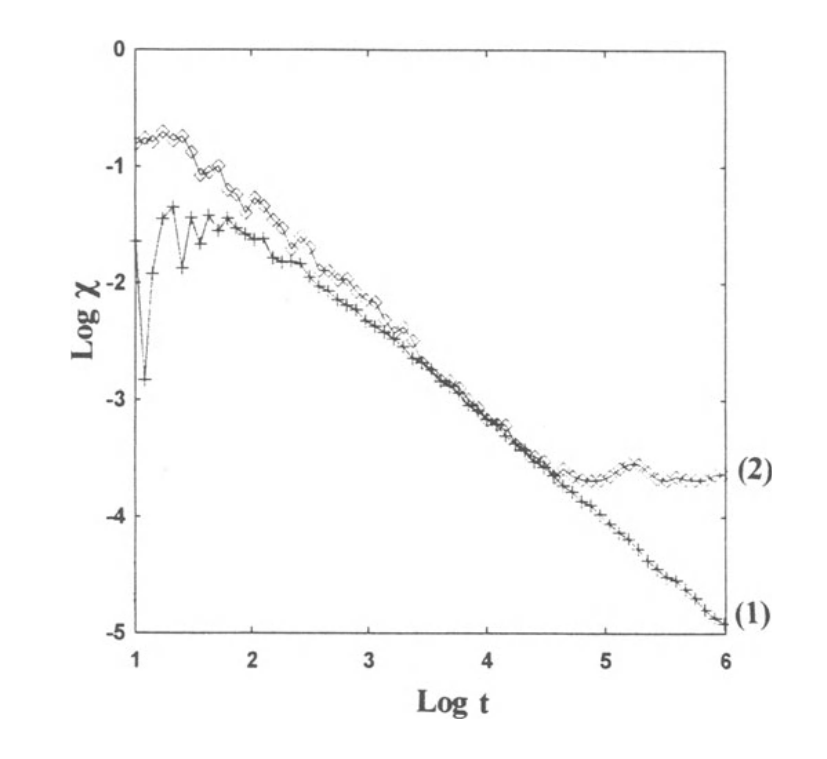}
\caption{The logarithm of the finite time LCN as a function of the logarithm of time in the cases of an ordered orbit (1) and a chaotic orbit (2).}
\label{fig9}
\end{figure}

However much more information is provided by tha ``local LCN'' or ``stretching number'':
\begin{align}
a_i=\ln(\xi_{i+1}/\xi{i})
\end{align}
where $\xi_i$ is the ith deviation in the case of mappings, or the ith intersection of a surface of section in the cae of Hamiltonian systems (Voglis and Contopoulos 1994). The distribution of the values of the stretching numbers define the spectrum of the dynamical system (Fig. 10).

\begin{figure}[H]
\centering
\includegraphics[scale=0.5]{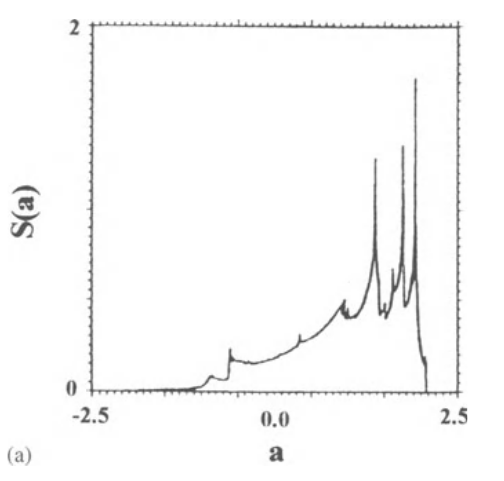}
\caption{The spectrum $S(a)$ of the stretching numbers $a$.}
\label{fig10}
\end{figure}

Much work has been done until recently on various aspects of chaos. While the dynamics of integrable and ergodic (in particular Anosov) systems are relatively simple the dynamics of general chaotic systems, where order and chaos coexist is much more difficult and complicated. The problems become more complicated if we consider 3-d systems and even more if we consider systems with more dimensions.

It is not possible to give an overview of the advances made on chaos up to now. An extensive review of these topics is provided by the book ``Order and Chaos in Dynamical Astronomy'' (Contopoulos 2002), I should mention also the book of Lichenberg and Liebermann `` Regular and Chaotic Dynamics'' (1992) and the book of Gutzwiller ``Chaos in Classical and Quantum Mechanics'' (1990). In the present review i want to concentrate on two important topics that developed in recent years (after 2006) namely a) An analytical study of chaos near unstable orbits and b) chaos in Bohmian Quantum Mechanics.

\section{2. Analytical Study of Chaos}

The integrals of motion around stable orbits are in general only formal. In fact the integrals of the general form (4) contain higher order terms with denominatoris of the form ($m_1\omega_1+m_2\omega_2$), where $m_1, m_2$ are positive or negative integers, while the ratio $\omega_1/\omega_2$ is irrational, and for convenient values of $m_1, m_2$ the denominator ($m_1\omega_1+m_2\omega_2$) can be arbitrarily small. These are the small divisors that make the corresponding terms of the integral $\Phi$ very large, thus producing non-convergence of the integral $\Phi.$ 

However in  an unstable periodic orbit one frequency is imaginary $\omega_2=i\nu$ (hence the term $1/2\omega_2^2x^2$ becomes $-1/2\nu^2x^2$) and in this case the divisor $m_1\omega_1+im_2\nu$ never becomes very small and the the corresponding integral is convergent. This was remarked by Cherry (1926) but it was expored in detail by Moser (1956, 1958) and Giorgilli (2001). Because of that we call the series $\Phi$ in this case ``Moser series''.

Detailed studies of Moser series were done by da Silva Ritter et al (1987), by Bongini et al. 2001, and by our group (Efthymiopoulos and Contopoulos 2006, Contopoulos and Efthymiopoulos 2008, Efthymiopoulos et al. 2007, 2009, 2014, Harsoula et al. 2015, Contopoulos and Harsoula 2015, Contopoulos et al. 2016, Harsoula et al. 2016, Contopoulos and Paez 2018). We studied in particular the domain of convergence of the Moser series and cases with more than one unstable periodic orbits.

We start with a case considered by da Silva Ritter (1987),  namely the hyperbolic Hénon map
\begin{align}
\nonumber&x'=\cosh(\kappa)x+\sinh(\kappa)y-\frac{1}{\sqrt{2}}\sinh(\kappa)x^2\\&
y'=\sinh(\kappa)x+\cosh(\kappa)y-\frac{1}{\sqrt{2}}\sinh(\kappa)x^2
\end{align}
Then we introduce variables

\begin{align}
u=(x+y)/\sqrt{2},\quad
v=(x-y)/\sqrt{2}\label{uva}
\end{align}
and 
\begin{align}
&\nonumber u=\xi+\Phi_{1}(\xi,\eta)\\&
v=\eta+\Phi_{2}(\xi,\eta)\label{uvb}
\end{align}
with
\begin{align}
\Phi_i=\Phi_{i,2}+\Phi_{i,3}+\dots
\end{align}
where $\Phi_{i,j} (i=1,2, j\geq2)$ are homogeneous polynomials in $(\xi, \eta)$, 
while $\xi$ and $\eta$ satisfy the simple relations 
\begin{align}
\xi'=\Lambda\xi,\quad \eta'=\eta/\Lambda
\end{align}\label{xieta}
with
\begin{align}
\Lambda=\lambda_1+w_{12}c+w_{13}c^2+\dots
\end{align}
where $w_j$ are constants and 
\begin{align}
c=\xi\eta=const
\end{align}
The series $\Phi_{i}$ ($i=1,2$) and $\Lambda$ converge within a region defined by 4 hyperbolas (Fig. 11)
\begin{align}
\xi\eta=\pm c_{crit}
\end{align}
For a given value of $c$ (where $|c|=c_{crit}$). and initial conditions ($\xi_0, \eta_0$) we find the images $(\xi_1, \eta_1)$ from the equations \ref{xieta} and the pre-images $(\xi_{-1}, \eta_{-1})$  by the inverse mapping and then the higher order images $(\xi_n, \eta_n)$ and pre-images $(\xi_{-n}, \eta_{-n})$. In practice we start at points $A_i (i=1, 2, 3, 4)$ where $\xi_0=\pm\eta_0=\pm\sqrt{c}$ and find the images $C_i$ and the pre-images $B_i$ (Fig. 11). Then we find the images of all the points of the arc $A_iC_i$ and the pre-images of the arc $A_iB_i$ on the plane $(x, y)$ using Eqs.12 and 11. An example of the curves $c=0$ and $c=\pm 1$ is given in Fig. 12. These curves are the stable (S, S') and unstable (U, U') asymptotic curves from the point $O(x=y=0)$ which in the plane ($\xi, \eta$) are represented by the axes $\xi$ and $\eta$. In a similar way we find the invariant curves for any given $c$ (where $|c|<c_{crit}$).

Then we find on the $(x, y )$ plane the convergence regions for all the values of $\xi, \eta$ that satisfy the conditions $|\xi\eta|\leq c_{crit}$ (Fig. 13).

The convergence region is quite simple on the plane $(\xi, \eta)$, its limits being defined by the hyperbolas $\xi\eta=\pm c_{crit}$. However on the plane ($x,y $) the convergence domain contains an infinity of lobes on the left and upwards or downwards plus two lobes on the right. In the case of Fig. 12 we have $K=1.43$. In this case there is a stable periodic orbit $s$. The convergence region does not come very close to $s$, but leaves an open domain (white in Fig. 13) around it.

\begin{figure}[H]
\centering
\includegraphics[scale=0.5]{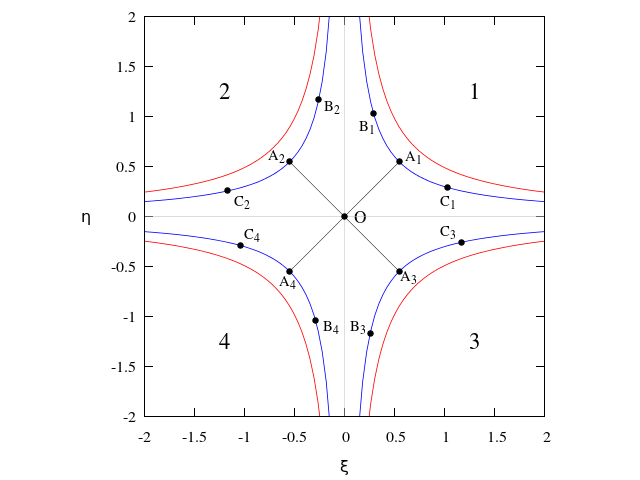}
\caption{The invariant hyperbolae on the plane $(\xi,\eta)$. We take the initial points $A_i,( i=1\dots 4)$ in the regions 1, 2, 3 and 4, along the diagonals $\xi=\pm\eta=\pm\sqrt{c}$. The images of  $A_i$ are $B_i$ and their preimages are $C_i$. The critical hyperbolae $(c=\pm 0.49)$ are red.}
\label{fig11}
\end{figure}

\begin{figure}[H]
\centering
\includegraphics[scale=0.5]{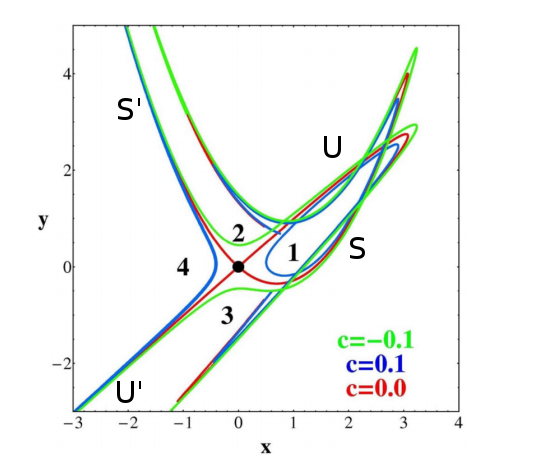}
\caption{The asymptotic curves $c=0$ from the central unstable orbit $(0,0)$ and the asymptotic curves $c=0.1$ from the regions 1 and 4 and $c=-0.1$ from the regions 2 and 3.}
\label{fig12}
\end{figure}

All the points in the convergence region have their images on the same convergence region and can be found by analytical formulae. Thus the chaotic domain around the unstable orbit $O(0,0)$ is foliated by invariant curves that are given by analytical formulae. However the successive points of the iteration (10) are further and further away from each other along any given invariant curve. E.g. for $\kappa=1.43$ we have $\lambda_1=4.18$, and after 5 iterations  we have $\lambda_1^5=1276$! However,  as the invariant curve has an infinity  of lobes we can see some high order images of an initial point in the neighbourhood of the origin (Fig. 14), although most images are at large distances along the lobes on the left upwards and downwards. In the same way the images of two nearby initial points with slightly different $c$ deviate considerably from each other after a few iterations.

Thus we have chaos, and this is not only deterministic (as it is given by formulae of the form 10) but furthermore it generates invariant curves through every point in the convergence region that can be described by analytical formulae.

Moreover we found that, if we start at points outside the convergence region, their images tend to approach the boundary of the convergence region. Namely in Fig. 15 we see images of the the points of the square $(-10, 10)\times(-10, 10)$ outside the convergence region approach closer the boundary of the convergence region at successive iterations. Of course these images can never enter into the convergence region. But the boundary of the convergence region acts as an attractor of the images of the points outside  the convergence region. Thus the images of the outer points after some iterations can be given approximately by the convergent series that define the boundary of the convergence region.
\begin{figure}[H]
\centering
\includegraphics[scale=0.5]{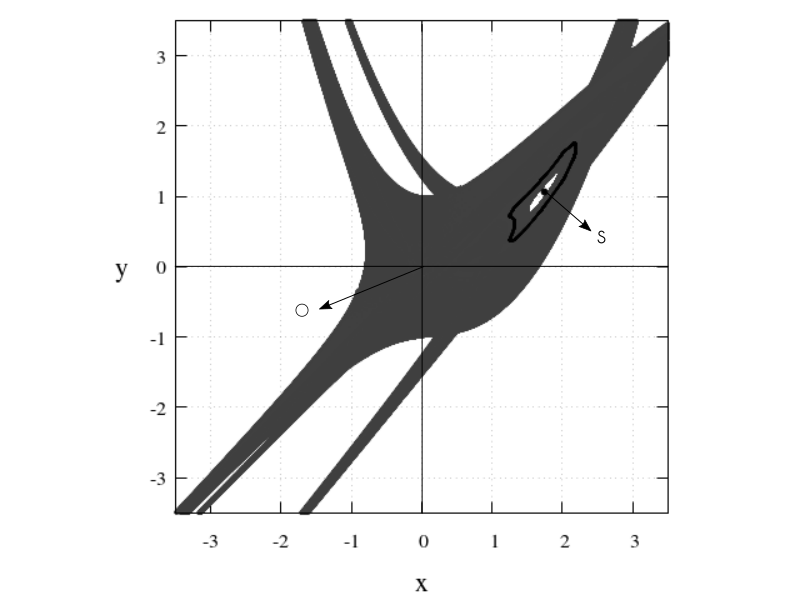}
\caption{The total convergence region around the origin O in the hyperbolic Henon map for $\kappa=1.43$. In this case there is a stable periodic orbit s, surrounded by a (dark) last KAM curve.}
\label{fig13}
\end{figure}
The up to now figures of this section refer to the case $\kappa=1.43$. But for a larger value of $k$ the orbit $s$ becomes unstable. In this case there is no empty region around $s$ and the convergence region around $O(0,0)$ reaches $s$.
On the other hand $s$ is now unstable and we can form Moser invariant curves around it (Fig. 16) by using a different set of variables $(\xi, \eta)$. In this case we have a different convergence region around $s$ (Fig. 16) that is contained completely inside the convergence region around $O$. Thus for the points inside the convergence region around $s$ we have two convergence formulae that give their images, one around $O$ and one around $s$.
\begin{figure}[H]
\centering
\includegraphics[scale=0.47]{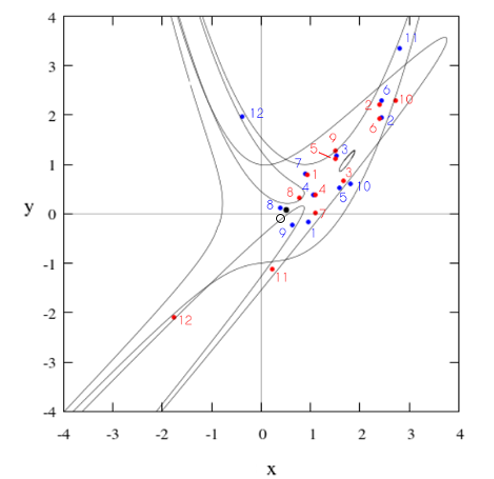}
\caption{The successive iterates of a point $o$ $(x=0.5064, y=0.08371)$ (red forward and blue  backward). All these points lie on the invariant curve $=0.3$. Superimposed are the boundaries of the convergence region $(x\pm 0.49)$. }
\label{fig14}
\end{figure}
In particular we can find analytically periodic orbits of higher order around $s$ that are generated from the orbit $s$ for values of $s$ for which $s$ is stable. We found also homoclinic orbits at the intersections of the asymptotic curves from O and at the intersections of the asymptotic orbits from $s$. Finally we found heteroclinic orbits at the intersections of the asymptotic curves from O with the asymptotic curves from $s$.

\begin{figure}[H]
\centering
\includegraphics[scale=0.47]{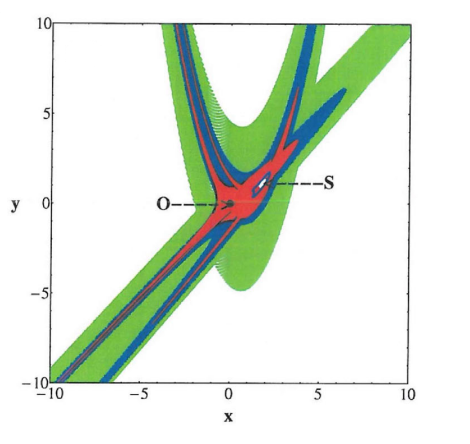}
\caption{The images of the points of the square $(-10,10)\times(-10,10)$ outside the convergence region (black) approach the boundary of the convergence region. The first iterations are yellow and the second iterations are orange.}
\label{fig15}
\end{figure}

Similar results are found for other mappings. In particular we have found invariant curves given by convergent series in the standard map

\begin{align}
&\nonumber x'=x+y'\\&
y'=y+\frac{K}{2\pi}\sin(2\pi x)
\end{align}

If K is larger than a critical value $K_{crit}=0.97$ chaos is dominant. We can consider this chaos as generated by the unstable periodic orbit $(O(x=y=0))$. We have studied in particular the case $K=2.7$.

We apply a transformation similar to (12),
such that $\xi'$ and $\eta'$ satisfy Eqs. (14), so that $\xi\eta=c$ is constant. The transformation (12) is convergent if $|c|$ is smaller than a critical value $c_{crit}=4.5$. In Fig. 17 we have drawn the asymptotic curves that correspond to $c=0$ (i.e. the images of the axes $\xi$ and $\eta$) in the case $K=2.7$. We have drawn also the invariant curves correspondig to $c=3.0$ and the invariant curves for $c=\pm 4$ that are close to the boundary of the convergence region $c_{crit}=4.5$. The upper line is the image of the hyperbola $\xi\eta=4$ with $\xi>0, \eta>0$ (region 1) in Fig. 10 and the lower line is the image of the hyperbola $\xi\eta=4$ with $\xi<0, \eta <0$ (region 4). The inner curves represent the hyperbolas $\xi\eta=-4$ (region 2 on th left and region 3 on the right). These curves are around the stable periodic orbits $s'$ and $s$.
\begin{figure}[H]
\centering
\includegraphics[scale=0.5]{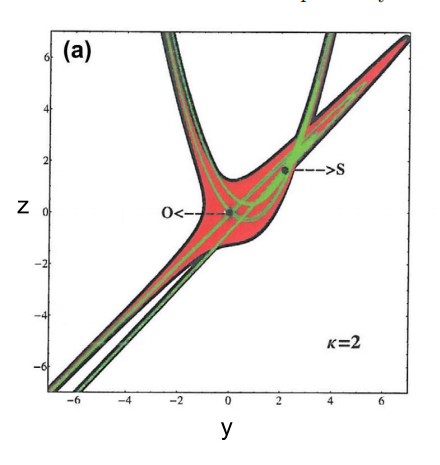}
\caption{In the case $\kappa=2.0$ the convergence region around the origin O is red and around s is green. In this case s is unstable. The green region is completely inside the red.}
\label{fig16}
\end{figure}

The orbits of initial conditions in the convergence region have images that diffuse to large values of positive $x$ and $\pm y$ and pre-images that diffuse to large negative $x$ and $\pm y$.
We calculate the diffusion parameter D in the equation (Harsoula et al. 2016)
\begin{align}
\langle (y-y_0)^2\rangle=Dn,
\end{align}
where n is the number of iterations for large n. In the case $K=2.7$ we have $D=0.05$.

\begin{figure}[H]
\centering
\includegraphics[scale=0.5]{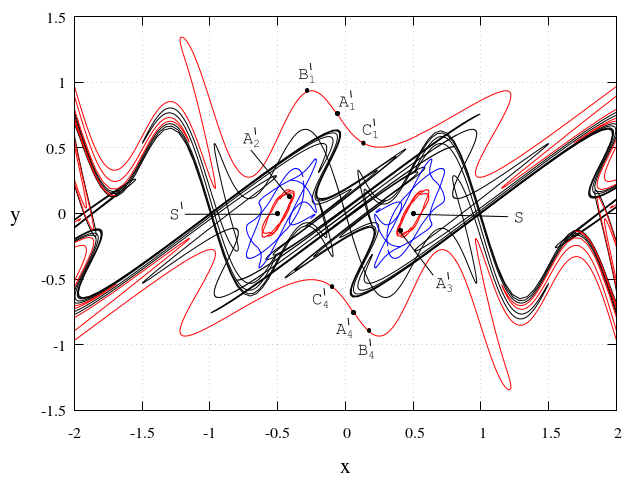}
\caption{The asymptotic curves from the unstable periodic orbit $O(0,0)$ in the standard map for  $K=2.7$. Superimposed are the limits of the convergence region in the regions 1 (upper curve), 4 (lower curve) and in the regions 2 and 3 around the stable periodic orbits s' and s.}
\label{fig17}
\end{figure}

In this case  we have also an approach of the images of points starting outside the convergence region to the boundary of the convergence region (Fig. 18). Thus in this case also we provide analytical solutions for the invariant curves and the images of points both inside and outside the convergence region. Although we have chaos we have a foliation of the chaotic domain that is given by analytical formulae.

\begin{figure}[H]
\centering
\includegraphics[scale=0.5]{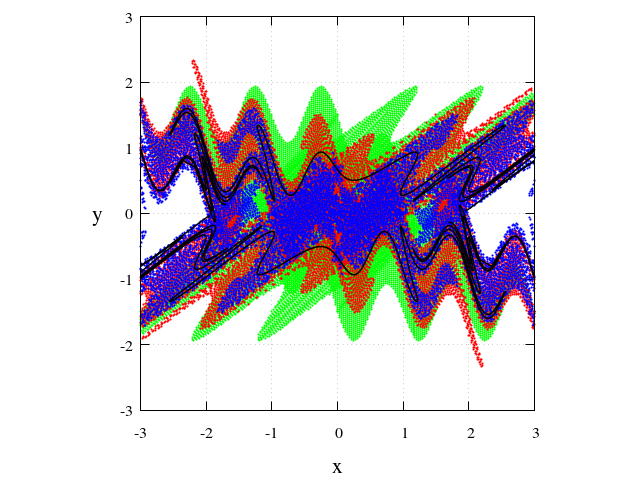}
\caption{The images of the orbits in the square $(-1.5, 1.5)\times(-1.5, 1.5)$ (both inside and outside the convergence region 1 approach the convergence region, at successive iterations (1st iteration green, 2nd red, 3rd blue). The limits of the convergence region are black.}
\label{fig18}
\end{figure}

Similar results are found for large values of $K$, where the orbit $s$ is unstable and we do not have a stable region round it, as in the case $K=6.0$ (Fig. 19). In the case $K=6.0$ we have a diffusion parameter $D=0.63$ much larger than in the case $K=2.7$.

Further cases where we have Moser series were provided for Hamiltonians (Efthymiopoulos et al 2014).

\begin{figure}[H]
\centering
\includegraphics[scale=0.5]{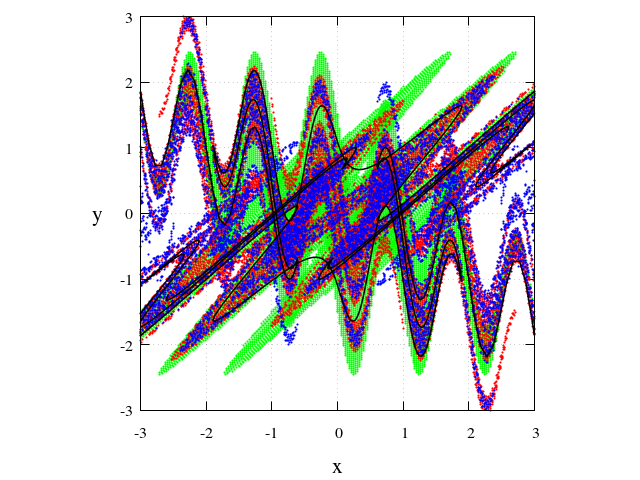}
\caption{As in Fig. 18 for $K=6.0$. In this case the orbits s' and s are unstable.}
\label{fig19}
\end{figure}

\section{3. Quantum Chaos}

Many people have studied chaos in quantum mechanics. However the Cophenhagen approach to this problem avoids the use of orbits, because, due to the Heisenberg uncertainty principle orbits are not well defined. Thus people study the behaviour of quantum systems when the corresponding classical systems are either ordered (integrable) or chaotic (containing in general both ordered and chaotic orbits).

\begin{figure}[b]
\centering
\includegraphics[scale=0.26]{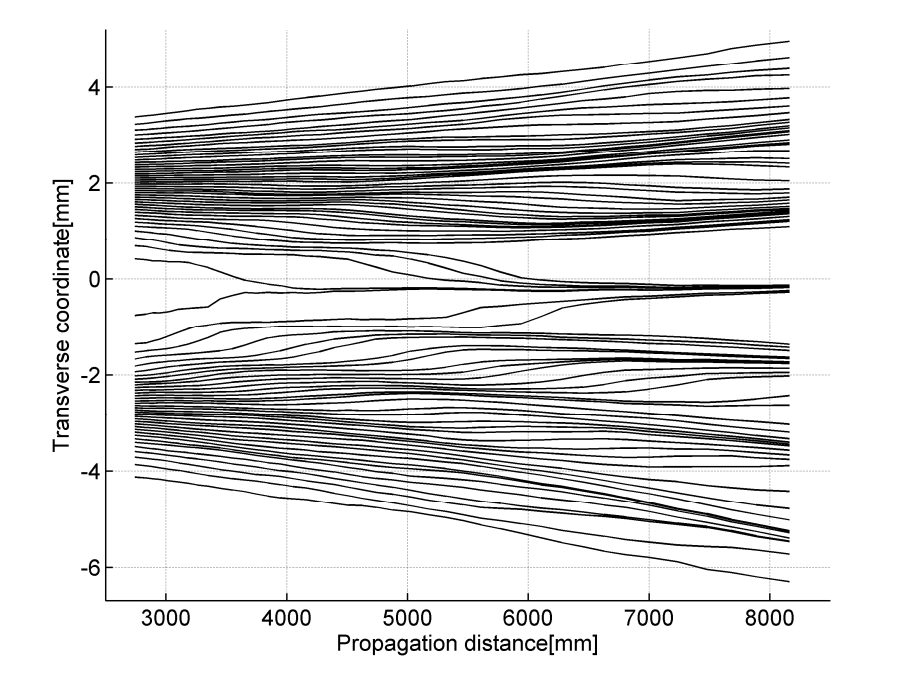}
\caption{Experimental representation of the Bohmian trajectories (Kocsis et al. 2011).}
\label{kocsis}
\end{figure}

On the other hand, the Bohmian approach to quantum mechanics (Bohm 1952) deals explicitly with orbits. It is remarkable that there are classically
ordered systems that are chaotic in the Bohmian approach of quantum mechanics and classically chaotic systems that are ordered in the Bohmian
approach (\cite{efthymiopoulos2006chaos}). Thus it is of interest to study quantum orbits for their own benefit. In recent years there have been experiments that describe the Bohmian orbits in the two-slit case (\cite{kocsis2011observing}) (Fig~\ref{kocsis}).
In order to avoid the Heisenberg uncertainty one has to make weak measurements of large numbers of similar orbits that form patterns very similar to the Bohmian orbits of individual particles.

The Bohmian orbits are solutions of the equations of motion: 
\begin{align}
\frac{d\vec{r}}{dt}=\Im\Big(\frac{\nabla\Psi}{\Psi}\Big)
\end{align}

Using this equation we can apply the methods of Dynamical Astronomy to study the orbits. In particular we use the Lyapunov characteristic number to find whether an orbit is ordered or chaotic.

\begin{figure}
\centering
\includegraphics[scale=0.4]{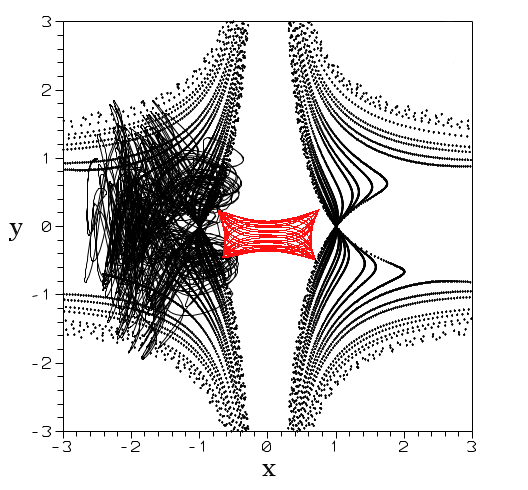}
\caption{The orbits of a nodal point (gray) and two orbits of particles, one ordered around the origin (0,0) (red) and one chaotic on the left (black).}
\label{mirella}
\end{figure}

When $\Psi=0$ (i.e. the real and the imaginary parts of the wavefunction are both zero), we have a nodal point of the system. The nodal point $(x_0, y_0)$ in general moves in time. Chaos is generated when an orbit approaches the region around the nodal point. E.g. in Fig.\ref{mirella} we see the paths of the nodal point and of two orbits. One is ordered like a Lissajous figure, that does not ever approach the nodal point, while the other is chaotic, approaching the nodal point from time to time (\cite{harsoulabohmian}).

\begin{figure}[H]
\centering
\includegraphics[scale=0.37]{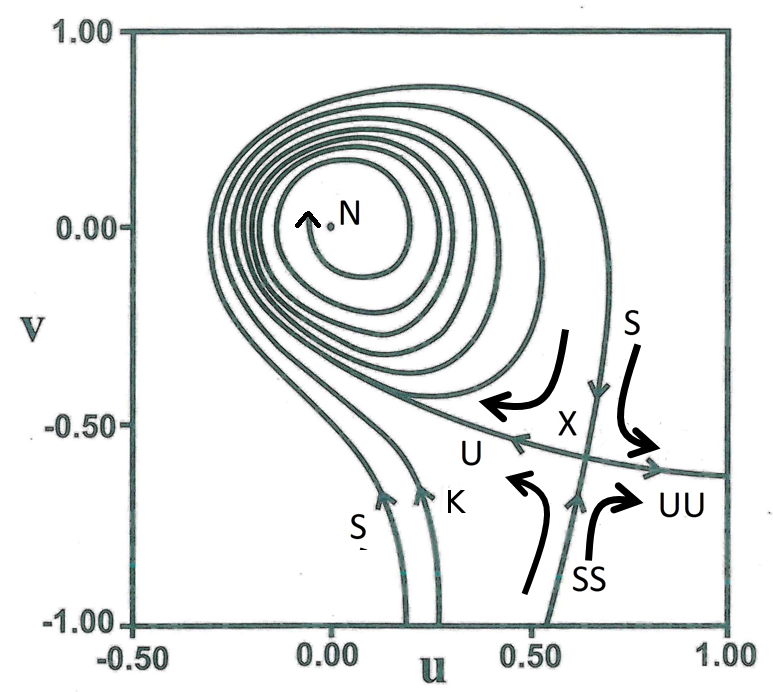}
\caption{The asymptotic curves of the unstable equilibrium X in the neighbourhood of a nodal point N, which in this case is an attractor. The unstable asymptotic curve U approaches asymptotically the nodal point N while the curve UU goes far away. The stable asymptotic curve S surrounds the region around N, before reaching X, while the curve SS reaches X directly. As regards non asymptotic orbits only those starting in the lower part of the figure like K between S and SS approach the nodal point N.}
\label{npxpc}
\end{figure}

An orbit that approaches the nodal point is a spiral in a frame moving with the nodal point ($u=x-x_0, v=y-y_0$) (Fig.~\ref{npxpc}). In the neighbourhood of the nodal point there is an unstable equilibrium point X. The X-point has two stable directions  S and SS (opposite to each other) and two unstable directions U and UU (again opposite to each other), and the corresponding asymptotic curves. One asymptotic curve (stable or unstable) goes towards the nodal point. The asymptotic curves may reach the nodal point N (after infinite rotations) as in Fig. \ref{npxpc}, or it may reach a limit cycle surrounding the nodal point. A detailed discussion of the various cases is provided in \cite{tzemos2018origin}. Chaos is generated when the orbits approach the X-point and then deviate close to unstable asymptotic curves in opposite directions (Fig.~\ref{npxpc}) \cite{Efth2009}.

In three dimensions we have sets of nodal points that form nodal lines and sets  of X-points that form X-lines (\cite{tzemos2018origin}). Chaos is produced when an orbit approaches the X-line (Fig.~\ref{npxpc}).

\begin{figure}[H]
\centering
\includegraphics[scale=0.4]{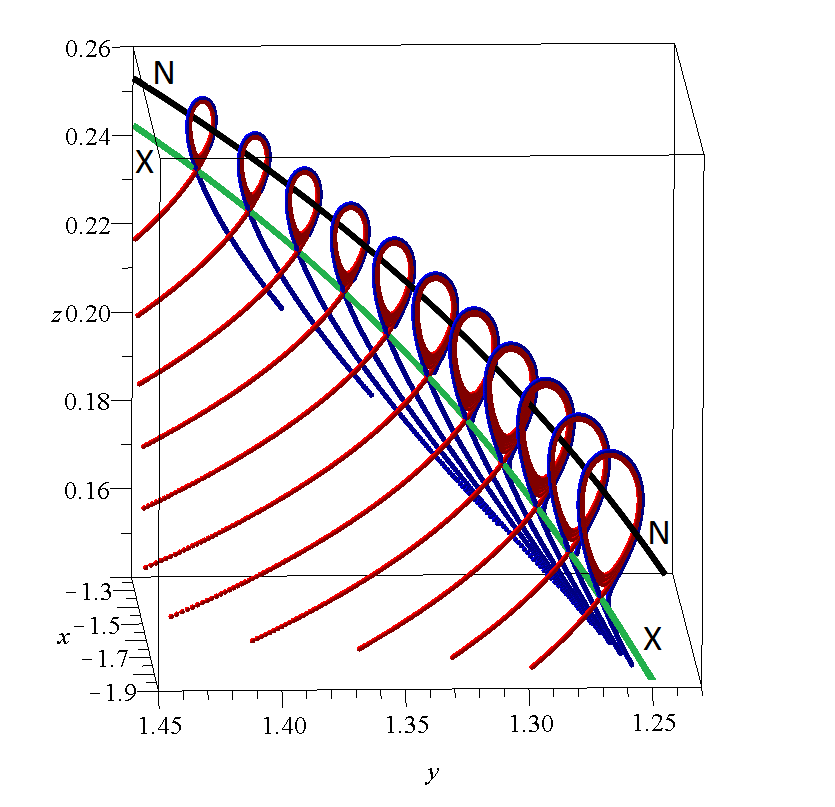}
\caption{In 3-d systems, asymptotic curves U starting at various points of an X-line marked X (green) form spirals (red) around nodal points on the nodal line (black curve marked N). The  unstable asymptotic curves (red) UU go downwards and to the left away from the nodal line. The stable curves (blue) SS go directly to points on the X-line, while the curves S (also blue) also start a little away from the SS lines (in fact the blue lines below X are double) and surround the spirals U counter-clockwise reaching  points on the X-line. }
\label{final}
\end{figure}

\begin{figure}[H]
\centering
\includegraphics[scale=0.33]{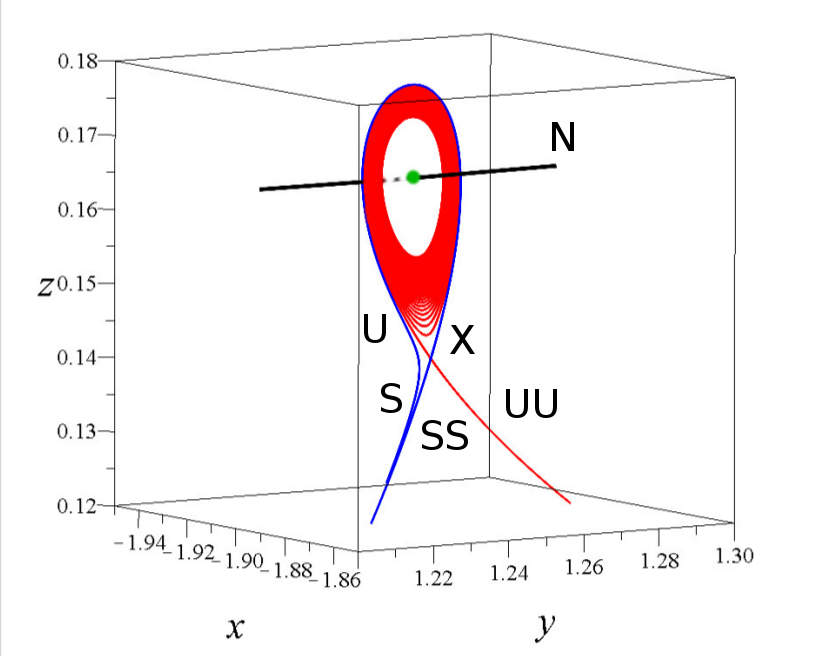}
\caption{A particular set of asymptotic orbits U, UU, S and SS from a particular point of the X-line and the nodal line N.}
\label{finalsingle}
\end{figure}

\begin{figure}[H]
\centering
\includegraphics[scale=0.38]{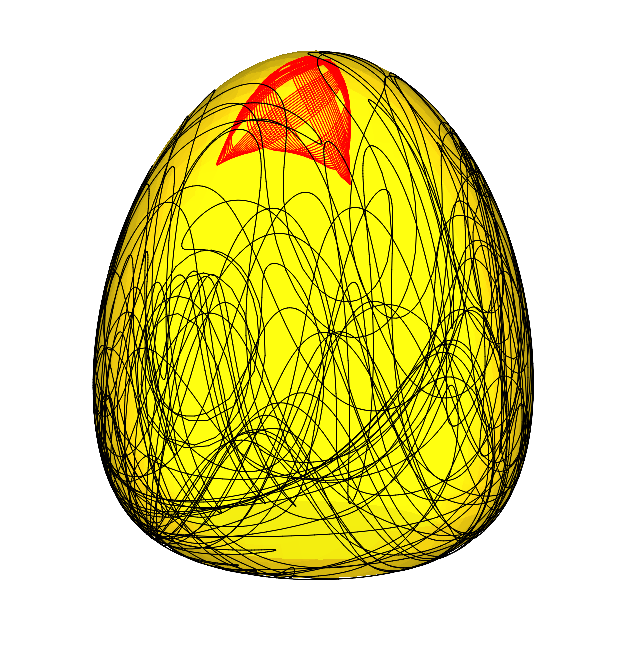}
\caption{The 3-d orbits of a quantum system of harmonic oscillators with wavefunction $\Psi=\frac{1}{\sqrt{3}}(\Psi_{100}+\Psi_{010}+\Psi_{002})$ lie on pear shaped surfaces. We give an ordered (red) orbit and a chaotic (black) orbit  on such a surface.}
\label{axladi}
\end{figure}

In 3 dimensions we have partially integrable cases, i.e. cases where the equations of motion satisfy an integral of motion (\cite {Tzemos2016}, \cite{contopoulos2017partial}, \cite{tzemos2018origin} and \cite{tzemos2018integrals}). E.g. in the case of 3d harmonic oscillators for particular wavefunctions $\Psi$ we have integral surfaces like the one of  Fig.~\ref{axladi}. Orbits starting on this surface remain always on in. Some orbits are ordered while others are chaotic. Only in very special cases we have complete integrability and no chaos at all.

However in more general cases of wavefunctions $\Psi$ the orbits are in general chaotic filling regions of 3 dimensions (complete nonintegrability).

Therefore as regards the orbits in quantum mechanics have all the properties of generic dynamical systems. Order and chaos, complete integrability, partial integrability and complete nonintegrability. The Bohmian approach to quantum mechanics opens an area of novel and very active research.

\begin{figure}[H]
\centering
\includegraphics[scale=0.47]{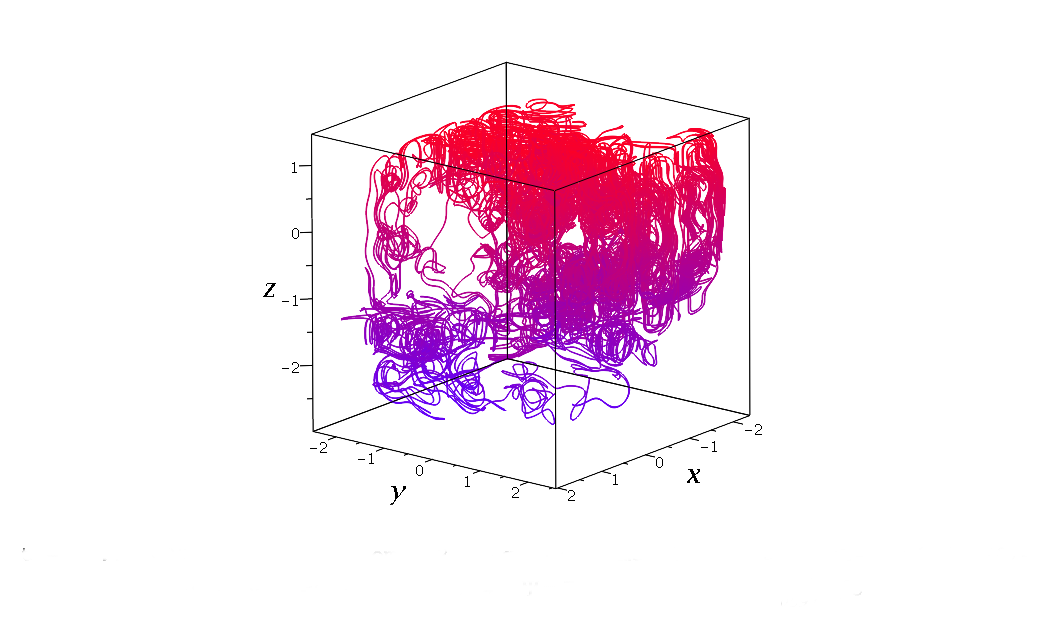}
\caption{The general form of 3-d chaotic orbits that do not lie on an integral surface.}
\label{fc}
\end{figure}

\nocite{*}
\bibliographystyle{agsm}
\bibliography{nbib}

\end{document}